\documentclass{iucrjournals}

\usepackage{graphicx}

\title{Laboratory Three-dimensional X-ray Micro-beam Laue Diffraction}

\author[1]{Yubin Zhang\IUCrCemaillink{yubz@dtu.dk}\IUCrOrcidlink{0000-0002-6901-0309}}%
\author[2]{Anthony Seret}%
\author[3]{Jette Oddershede}%
\author[3]{Azat Slyamov}
\author[4]{Jan Kehres}
\author[3]{Florian Bachmann}
\author[4]{Carsten Gundlach}
\author[4]{Ulrik Lund Olsen}
\author[3]{Jacob Bowen}
\author[4]{Henning Friis Poulsen}
\author[3]{Erik Lauridsen}
\author[1]{Dorte {Juul Jensen}}

\affil[1]{Technical University of Denmark, Department of Civil and Mechanical Engineering, 2800 Kgs. Lyngby, Denmark}
\affil[2]{Boulevard de l’Observatoire, CS 34229 - F 06304 NICE Cedex 4, France}
\affil[3]{Xnovo Technology ApS, 4600 Køge, Denmark}
\affil[4]{Technical University of Denmark, Department of Physics, 2800 Kgs. Lyngby, Denmark}

\begin{document} 
\maketitle 

\begin{synopsis}
A novel laboratory-based 3D X-ray micro-beam diffraction (Lab-3D$\mu$XRD) technique has been developed and successfully validated. The setup enables the detection of grains as small as 10~$\mu$m, with an intragranular orientation uncertainty of 0.01$^\circ$.
\end{synopsis}

\begin{abstract}
The development of three-dimensional (3D) non-destructive X-ray characterization techniques in home laboratories is essential for enabling many more researchers to perform 3D characterization daily, overcoming the limitations imposed by competitive and scarce access to synchrotron facilities. Recent efforts have focused on techniques such as laboratory diffraction contrast tomography (LabDCT), which allows 3D characterization of recrystallized grains with sizes larger than 15–20 $\mu$m, offering a boundary resolution of approximately 5$\mu$m using commercial X-ray computed tomography (CT) systems. To enhance the capabilities of laboratory instruments, we have developed a new laboratory-based 3D X-ray micro-beam diffraction (Lab-3D$\mu$XRD) technique. Lab-3D$\mu$XRD combines the use of a focused polychromatic beam with a scanning-tomographic data acquisition routine to enable depth-resolved crystallographic orientation characterization. This work presents the first realization of Lab-3D$\mu$XRD, including hardware development through the integration of a newly developed Pt-coated twin paraboloidal capillary X-ray focusing optics into a conventional X-ray $\mu$CT system, as well as the development of data acquisition and processing software. 
The results are validated through comparisons with LabDCT and synchrotron phase contrast tomography. The findings clearly demonstrate the feasibility of Lab-3D$\mu$XRD, particularly in detecting smaller grains and providing intragranular information. Finally, we discuss future directions for developing Lab-3D$\mu$XRD into a versatile tool for studying materials with smaller grain sizes and high defect densities, including the potential of combining it with LabDCT and $\mu$CT for multiscale and multimodal microstructural characterization.
\end{abstract}

\keywords{three dimensional X-ray diffraction (3DXRD), laboratory diffraction contrast tomography, Laue micro-beam diffraction, scanning 3DXRD, intragranular orientation}

\section{Introduction}

Advances in the characterization of crystalline materials are essential for unraveling their processing-structure-property relationships, guiding the development of next-generation materials with tailored properties. Among the many techniques available, three-dimensional (3D) non-destructive grain mapping using X-rays has emerged as a cornerstone for gaining detailed insights into the internal microstructure, local crystallographic orientations, and defect distributions of materials \cite{JuulJensen2000, Offerman2002, King2008, Barabash2009, Ice2011, JuulJensen2012, Pokharel2014, JuulJensen2020, Shahani2020}. Broadly, these techniques can be categorized into two families: those using monochromatic beams, such as 3D X-ray diffraction (3DXRD) \cite{Poulsen2004,Johnson2008, Li2012}, and those using polychromatic beams, such as Laue 3D micro-beam X-ray diffraction (3D$\mu$XRD) \cite{Larson2002}. 

3DXRD typically employs a monochromatic beam and a tomographic data acquisition to map grain orientations and strain fields in 3D. With a line or a box beam, a large portion of, or even an entire grain is illuminated by the beam, limiting the method’s capability to resolve intragranular information \cite{Li2012, Pokharel2015, Vigano2016, Winther2017}. To improve spatial resolution, focused beam configurations and scanning methods have been introduced, leading to the development of scanning 3DXRD (S3DXRD) \cite{Hayashi2019, Wright2020, Li2023, Henningsson2024}. 3D$\mu$XRD uses a focused polychromatic beam combined with differential aperture scanning to achieve depth-resolved orientation indexing, avoiding the need for sample rotation \cite{Larson2013}. After 20 years of development, these techniques have become indispensable tools in materials science. Nonetheless, synchrotron-based methods are inherently limited by the accessibility of large-scale facilities—extended measurement times (\textgreater 1 week) are not realistic, and the turn-around time (from idea to result) is very long, typically a year or more.

To address these limitations, significant efforts have been directed toward adapting advanced X-ray characterization methods in home laboratories. One such method, laboratory X-ray diffraction contrast tomography (LabDCT), has become an established 3D characterization tool \cite{King2013LabDCT, Feser2015, McDonald2015, Holzner2016,  Oddershede2022}. LabDCT utilizes a conical polychromatic X-ray beam generated from laboratory X-ray tubes in conventional X-ray CT systems, combined with diffraction principle based on the Laue focusing effect \cite{Kvardakov1997, Guinier1949}. It enables the 3D mapping of grain orientations and morphologies without requiring access to synchrotron facilities. However, even by using all the flux from the entire polychromatic X-ray spectrum from laboratory X-ray tubes, it remains challenging to use this technique to map grains smaller than 20 $\mu$m \cite{Fang2021, Fang2023}. Like its synchrotron predecessor, DCT—a variation of 3DXRD—the use of a large beam limits its application for studying deformed materials with intragranular orientation or strain variations.

At the same time, laboratory micro-beam diffraction (Lab-$\mu$XRD) setups have been developed, enabling the characterization of intragranular information in thin samples \cite{Lynch2007, Lynch2019, Zhang2024}. However, so far, this technique has only been used for 2D mapping. This is mainly because of the weak beam intensity in laboratory sources, which makes techniques like the differential aperture method \cite{Larson2002}—commonly used at synchrotrons for depth-resolved (the third dimension) orientation indexing—not appropriate in laboratory settings.

To enable high resolution 3D mapping in the laboratory, we introduce a new method that combines the principles of S3DXRD and Lab-$\mu$XRD: laboratory 3D micro-beam X-ray diffraction (Lab-3D$\mu$XRD). Specifically, we propose to combine the use of a focused polychromatic beam (inspired from $\mu$XRD) and the scanning-tomographic data acquisition routine of S3DXRD, to enable 3D non-destructive characterization of crystalline materials in a home laboratory environment \cite{Zhang2020}. This article aims to demonstrate the proposed concept, including i) the principles underlying the method, ii) its first experimental implementation (encompassing both hardware and data acquisition software), iii) the development of software for processing the collected data, and iv) its validation through comparative studies with LabDCT and synchrotron phase contrast tomography (PCT). We demonstrate the feasibility of this approach for high spatial resolution characterization. Additionally, we discuss possible future directions for developing the technique into a versatile tool for materials studies, including its potential combination with LabDCT and conventional $\mu$CT to enable multiscale and multimodal analysis.

\begin{figure}[ht] %
\label{fig:lab3dxrd}
\begin{center}
\includegraphics[width=0.85\textwidth]{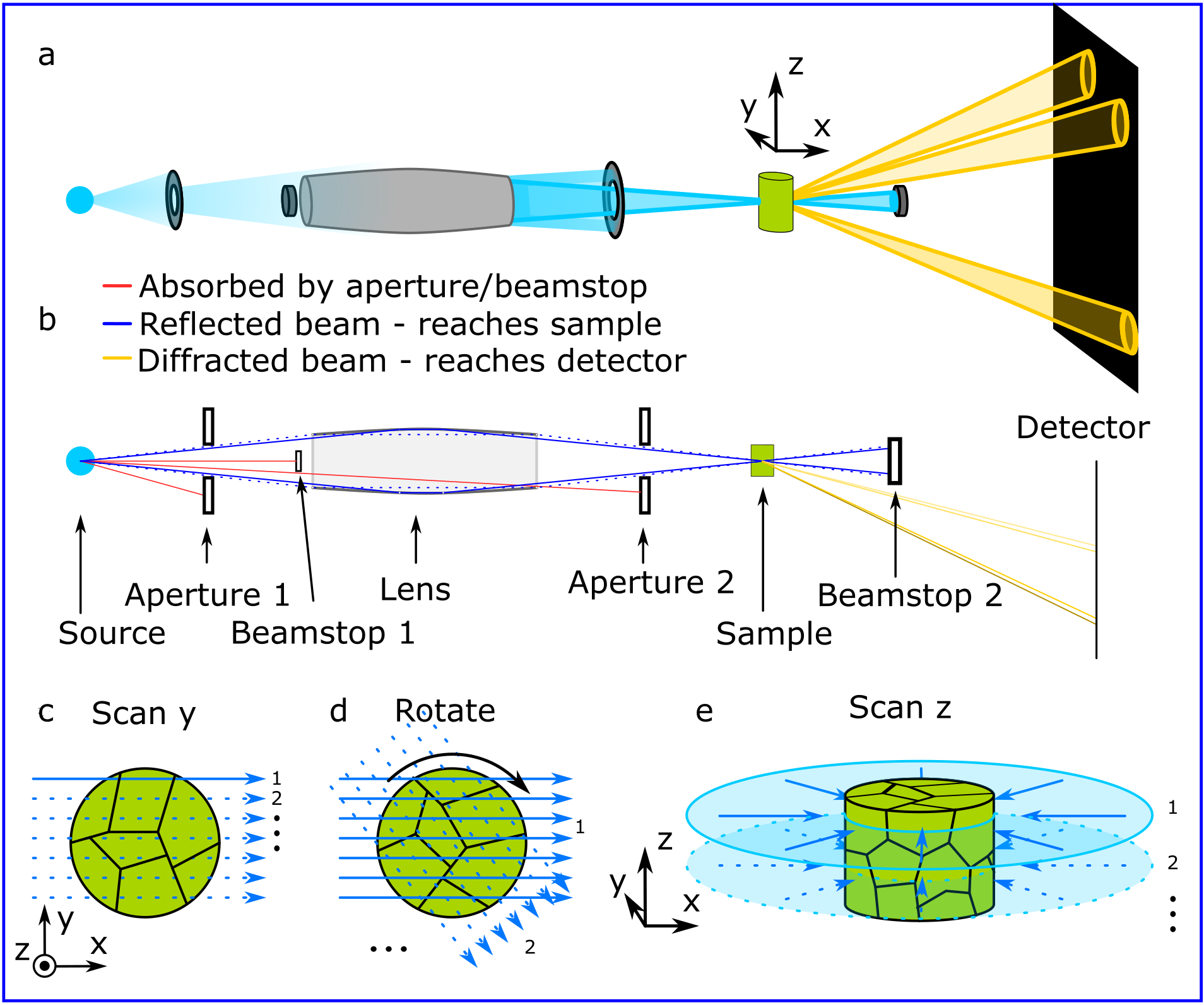} 
\end{center}
\caption{Schematic illustration of the Lab-3D$\mu$XRD geometry (a and b) and the data acquisition routine (c-e). The dashed and solid blue lines in (b) represent example incident X-rays with lower and higher energies, respectively. The yellow lines represent diffracted X-rays, with variations in brightness indicating changes in wavelength—brighter lines correspond to shorter wavelengths (i.e., higher energies). The angles between the lines, i.e. the beam divergence, are significantly exaggerated. Arrows in (c) and (d) indicate  the beam paths into the sample during its translation.}
\end{figure}

\section{Principles of Lab-3D\texorpdfstring{$\mu$}{mu}XRD}

The principles of Lab-3D$\mu$XRD are sketched in Fig.~\ref{fig:lab3dxrd}. Similar to LabDCT, a polychromatic X-ray beam is employed to maximize the utilization of the available photons. An X-ray source with a heavy element target, such as tungsten, is preferred over conventional Cu or Cr sources because the intensity of emitted X-rays increases approximately linearly with the atomic number of the target. In addition, focusing X-ray optics are introduced to further enhance intensity. These optics not only amplify the X-ray flux by concentrating the beam—beneficial for detecting smaller grains—but also limit the field of view illuminated within the sample, which is critical for enabling intragranular characterization. For the characterization of metals such as Al, Ti, and steel, it is crucial that the optics can focus high-energy X-rays, ensuring that submillimeter- to millimeter-scale samples can be effectively studied \cite{Seret2023}. 

As with S3DXRD and synchrotron microdiffraction techniques, a detector with a relatively large pixel size is suitable for Lab-3D$\mu$XRD. The detector should be positioned in transmission geometry rather than in 90$^\circ$ reflection or 180$^\circ$ back-projection mode. This placement simplifies implementation and maximizes the Lorentz-polarization factor \cite{Warren1990}, which is critical for enhancing the intensity of the diffraction signals. Similar to LabDCT, a beamstop is used to block the transmitted beam, preventing detector saturation and enhancing the detection of the weaker diffraction signals, see Fig.~\ref{fig:lab3dxrd}(a) and (b).

With a focused beam, it is straightforward to confine the gauge volume to a narrow channel through the sample. However, achieving full 3D characterization requires recovering depth information, i.e., determining the precise location along the gauge channel where diffraction occurs. To address this challenge, we propose a new tomographic data acquisition method to restore depth information and retrieve full 3D crystallographic data, adapting the principles of S3DXRD as originally proposed in \citeasnoun{Poulsen2004}:

\begin{enumerate}
    \item Laue diffraction images are collected at every step while the sample is scanned along the horizontal direction, $y$ in Fig.~\ref{fig:lab3dxrd}(a and c), perpendicular to the incoming beam, $x$.
    \item The sample is rotated 360$^\circ$ around the vertical axis ($z$) at equal-angular intervals, and the scan along y is repeated after each rotation (see an example in Fig.~\ref{fig:lab3dxrd}(d)).
    \item By repeating the measurement layer by layer (see Fig.~\ref{fig:lab3dxrd}(e)), a 3D image of the grain structure can be obtained.
\end{enumerate}

In this way, intragranular information, such as the local crystallographic orientation and strain, in any volume element within the scanned section can be determined by combining the diffraction images collected as the beam passes through this location at different rotation angles \cite{Hayashi2015,Henningsson2020,henningsson2023}.

\section{Experimental Implementation}

\subsection{Material}
A beta titanium alloy (Ti-$\beta$21S) sample with a 300~$\mu$m diameter was chosen for this proof-of-concept experiment, as its grain structure was already known from previously high resolution synchrotron PCT measurement at beamline ID19 of the European Synchrotron Radiation Facility. The microstructure consists of equiaxed grains of the metastable $\beta$ phase with a body-centered cubic (bcc) lattice. The sample was annealed at 830~°C for 30~min to promote grain growth, yielding an average grain size of ~40~$\mu$m. A subsequent annealing at 725~°C for 15~min formed a thin hexagonal (hcp) $\alpha$ phase layer along the grain boundaries. The difference in chemical composition between the two phases providing a detectable electron density contrast, enabling 3D grain shape determination via PCT. The submicron-resolution (0.56~$\mu$m) 3D grain structure serves as an ideal benchmark for validating Lab-3D$\mu$XRD results. More details on the PCT experiment and grain segmentation can be found in \citeasnoun{McDonald2015}.

\subsection{Hardware development}

To demonstrate the concept, a conventional X-ray tomography instrument, Zeiss Xradia Versa 520, equipped with a flat panel extension (FPX) and a LabDCT Pro module, was utilized for the experiment. This system featured a tungsten-target microfocus X-ray source, capable of operating at a maximum energy of 160~keV and a power of 10~W. Pt-coated twin paraboloidal X-ray optics, designed and manufactured by Sigray Inc., were used to focus the X-ray beam. These optics were designed for maximum efficiency at a photon energy of approximately 32~keV. In practice, they are capable of focusing the X-ray beam to a full width at half maximum (FWHM) of 8.5~$\mu$m with a divergence angle of 0.59$^\circ$. The focal point was located 200~mm downstream from the X-ray source and 30~mm from the exit of the optics (i.e., working distance). The resulting focused beam was achromatic, supporting a broad energy range of up to 45~keV. Further details on the characteristics of the optics can be found in \citeasnoun{Seret2023}.

For precise alignment of the optics, a manually adjustable five-axis stage from Thorlabs was employed. To minimize stray radiation, a 400~$\mu$m upstream aperture was used, along with an additional 2~mm aperture located downstream of the optics.  A Dexela 2315 flat panel detector, mounted in transmission geometry, was used to record Laue diffraction patterns. The scintillator-based CMOS detector featured a resolution of 1936$\times$3064 pixels and a physical pixel size of 75~$\mu$m. To block the direct beam and enhance the contrast of the diffracted signals, a 2~mm thick tungsten beamstop was positioned centrally on the detector. 

\subsection{Data acquisition}

For the Lab-3D$\mu$XRD experienment, an acceleration voltage of 110~kV and a source power of 10~W were used. These parameters were chosen to optimize the photon flux based on insights from previous studies \cite{Lindkvist2021}. Following a manual inspection of diffraction images collected at different sample-to-detector distances ($L_{sd}$), the flat panel detector was positioned at an $L_{sd}$ of 100~mm to ensure that most of the strong diffraction peaks were effectively captured. An exposure time of 105~s was used for each projection, with no detector binning applied.

During the data acquisition, the sample was scanned horizontally with a translation step size of 5~$\mu$m at each of the 36 rotation angles, spaced 10$^\circ$ apart. At each rotation angle, a reference image made by averaging five consecutive acquisitions was collected to normalize the diffraction images, ensuring consistent image quality across the dataset. The total scan time for this Lab-3D$\mu$XRD experiment was 86 hours. Data acquisition was performed using an in-house developed control framework, designed to orchestrate the imaging system components. For this initial demonstration, only a single slice near the center of the PCT volume was characterized.

To validate the characterized crystallographic orientations, the sample was also scanned using LabDCT with the same detector and mounting setup. The LabDCT measurement was conducted using the same X-ray power and acceleration voltage as Lab-3D$\mu$XRD. A total of 181 projections were collected with an exposure time of 60~s during a full 360$^\circ$ rotation, at a source-to-sample distance ($L_{ss}$) of 11.5~mm and an $L_{sd}$ of 245.8~mm, respectively, i.e., in a projection geometry \cite{Oddershede2022}. In addition, a $\mu$CT scan was performed to reconstruct the sample shape and facilitate the analysis of the LabDCT data. The total scan time for LabDCT and $\mu$CT was about 4 hours. 

\subsection{Processing of Lab-3D\texorpdfstring{$\mu$}{mu}XRD data}

The process consists of three main steps for each layer:  
\begin{enumerate}
    \item preprocessing raw diffraction images,  
    \item indexing grain orientations, and  
    \item reconstructing the grains and post-processing the 3D grain structure.  
\end{enumerate}

All three steps were performed in MATLAB, with each step detailed in the following subsections. For processing of data from multiple layers, these three steps will be repeated for each layer. However, the results from the first layer can serve as prior knowledge to accelerate the entire process.

\subsubsection{Preprocessing for spot detection}

\begin{figure}[ht]
    \centering
    \includegraphics[width=0.95\textwidth]{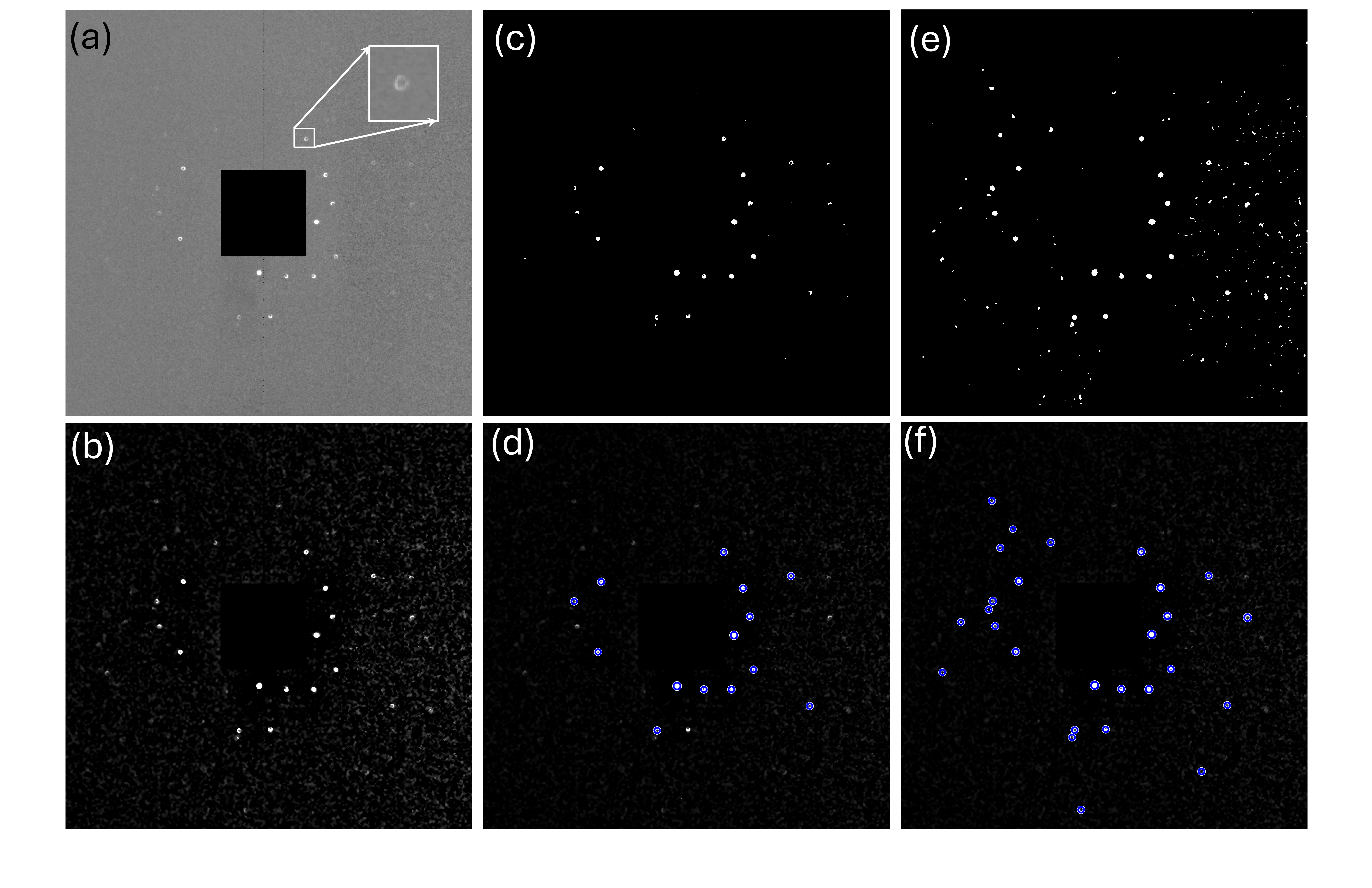} 
    \caption{An example of a diffraction image and the processing routine: (a) normalized diffraction image; (b) after background removal and image processing; (c) and (e) binarized images of (b) using higher and lower threshold values, respectively; (d) and (f) corresponding detected diffraction spots. Each image represents a physical size of 112.5~$\times$~112.5~$\mu$m$^2$. The central black region is due to the beam stop.}
    \label{fig:diffraction_spots}
\end{figure}

The collected raw diffraction images were processed in groups based on the sample rotation angle. For each group, the raw images were first normalized by dividing by the reference images from the same rotation angle \cite{Lindkvist2021}. An example of the central region of a normalized diffraction image with many doughnut-shaped diffraction spots, is shown in Fig.~\ref{fig:diffraction_spots}(a). The doughnut shape originates from the incoming beam, where the central rays do not interact with the paraboloidal mirrors and therefore are blocked by an interior small tungsten pillar, preventing them from passing through the optics unfocused \cite{Seret2023}, see Fig.~\ref{fig:lab3dxrd}(b). 

After normalization, intensity fluctuations between images may still be noticeable over time. These variations are caused by X-ray source fluctuations and were corrected using scaling factors determined from the average intensity over a small 100$\times$100 pixel area that does not contain visible diffraction spots. A background, determined as the median of the group images collected at each rotation angle, was then subtracted from each intensity-scaled image (see Fig.~\ref{fig:diffraction_spots}(b)). To further reduce long-range intensity variations, a Gaussian filter with a sigma of 50 pixels was applied to each of the background-corrected images, and the resulting Gaussian-filtered images were then subtracted from the originals. As a final step in image processing, a non-localized mean filter was applied to suppress any remaining local noise.

Different threshold values were then subsequently applied to segment diffraction spots, with spots smaller than 5 pixels eliminated to minimize noise effects. Binary images segmented using comparatively high and low threshold values are shown in Fig.~\ref{fig:diffraction_spots}(c) and (e), respectively. While the low threshold yielded a higher number of detected spots, it also resulted in some false detections. As to be described later, both thresholds were used in the analysis process. Finally, to improve spot center determination, a ring detection algorithm based on the Circular Hough Transform was employed to identify the doughnut-shaped spots (see Fig.~\ref{fig:diffraction_spots}(d) and (f)).

\subsubsection{Orientation indexing}

\begin{figure}[ht]
    \centering
    \includegraphics[width=0.85\textwidth]{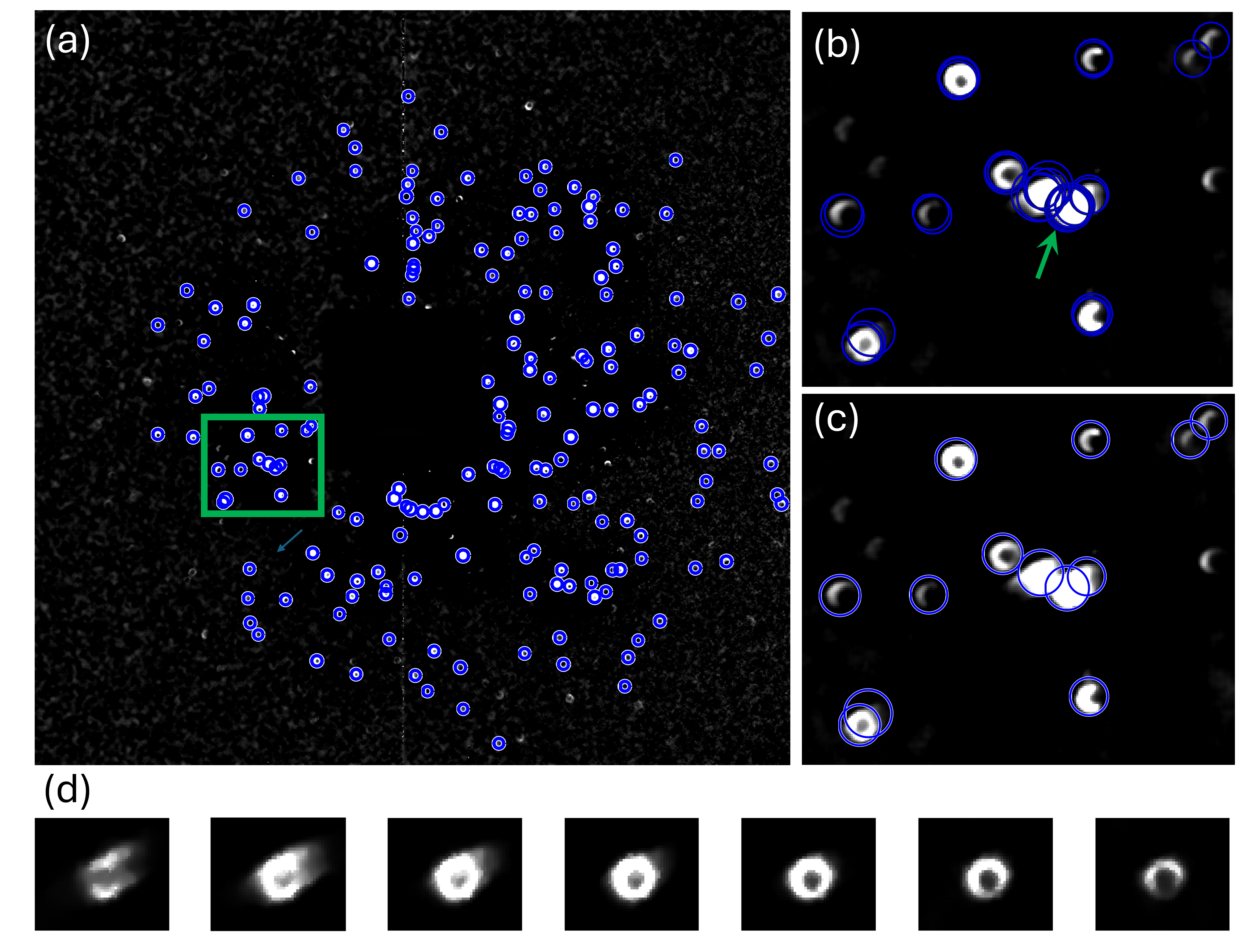} 
    \caption{(a) An example showing a set of diffraction spots, where those appearing at the same location on the detector are merged. (b) A magnified view of all detected diffraction spots from images collected at -180$^\circ$ rotation angle, overlaid onto a sum of all diffraction images, and (c) the corresponding merged spot set in the region marked by the green rectangle. (d) The diffraction spot, marked by the green arrow in (b), appearing at different translation steps.}
    \label{fig:merged_spots}
\end{figure}

The centers of the detected rings were used to calculate the diffraction vectors for indexing grain orientations. To facilitate the indexing, diffraction spots appearing at nearby pixel positions (within a 5-pixel distance in this work) across all individual diffraction images collected during sample translation at the same rotation angle were merged into a single spot. For the present case, this process resulted in 36 sets of diffraction spots corresponding to the 36 rotation steps. An example of the merged diffraction spots for the first rotation angle is shown in Fig.~\ref{fig:merged_spots}. This merging process significantly reduced the number of diffraction spots for orientation indexing, which was done using the laboratory dictionary-based branch-and-bound algorithm (LabDBB) \cite{Zhang2025}.

The LabDBB algorithm operates by comparing experimental diffraction vectors with theoretical diffraction vectors, calculated from a predefined orientation dictionary. Candidate orientations from the dictionary, known as orientation branches, are determined when a sufficient number of theoretical diffraction vectors from an orientation match experimental vectors within a specified upper-bound deviation angle, defined by the dictionary branch resolution. Once a candidate orientation is identified, it is refined to determine its validity based on additional threshold criteria, such as the number of matched diffraction vectors and their angular deviation. It should be noted in experiments using a polychromatic X-ray beam, the exact photon energies corresponding to the diffraction spots are not known. As a result, only normalized diffraction vectors are considered during orientation matching. This approach ensures accurate and efficient determination of grain orientations within the scanned sample layer. Further details about the LabDBB algorithm can be found in \citeasnoun{Seret2022} and \citeasnoun{Zhang2025}.

To accelerate the indexing process and improve accuracy, a two-step indexing approach was performed. In the first step, diffraction spots detected using the higher threshold were utilized to index the initial grain orientations. This ensured a focus on strong and clearly defined diffraction spots. The indexed orientations were then applied to fit the detector geometry using the method described in \citeasnoun{Fang2022}, after which they were further refined. In the second step, diffraction spots uniquely assigned to each indexed orientation from the higher threshold set were removed from the lower threshold set. This step effectively filtered the dataset by eliminating already-indexed spots, allowing the algorithm to focus on the additional diffraction spots detected with the lower threshold. The remaining unassigned diffraction spots were then used for further indexing, incorporating weaker diffraction signals into the analysis.

For the present case, a dictionary branch resolution of 2.5$^\circ$ was used, resulting in 39,565 orientations generated with MTEX \cite{Bachmann2011}. For each of the two intensity thresholds, a completeness threshold decreasing from 0.7 to 0.4, in steps of 0.1, was applied to limit the number of candidates requiring further time-consuming orientation fitting at each step. The final indexed orientations required a completeness value of 0.3, with their diffraction vectors deviating by no more than 0.25$^\circ$ from the corresponding theoretical predictions. The completeness value is defined as the ratio of the observed number of spots to the theoretical number of spots, calculated using the first three \{hkl\} families of a body-centered cubic $\beta$-Ti crystal structure and an X-ray energy range of 15–45~keV.

\subsubsection{Reconstruction of grains and post-processing of assembled grain structure}

\begin{figure}[ht]
    \centering
    \includegraphics[width=0.85\textwidth]{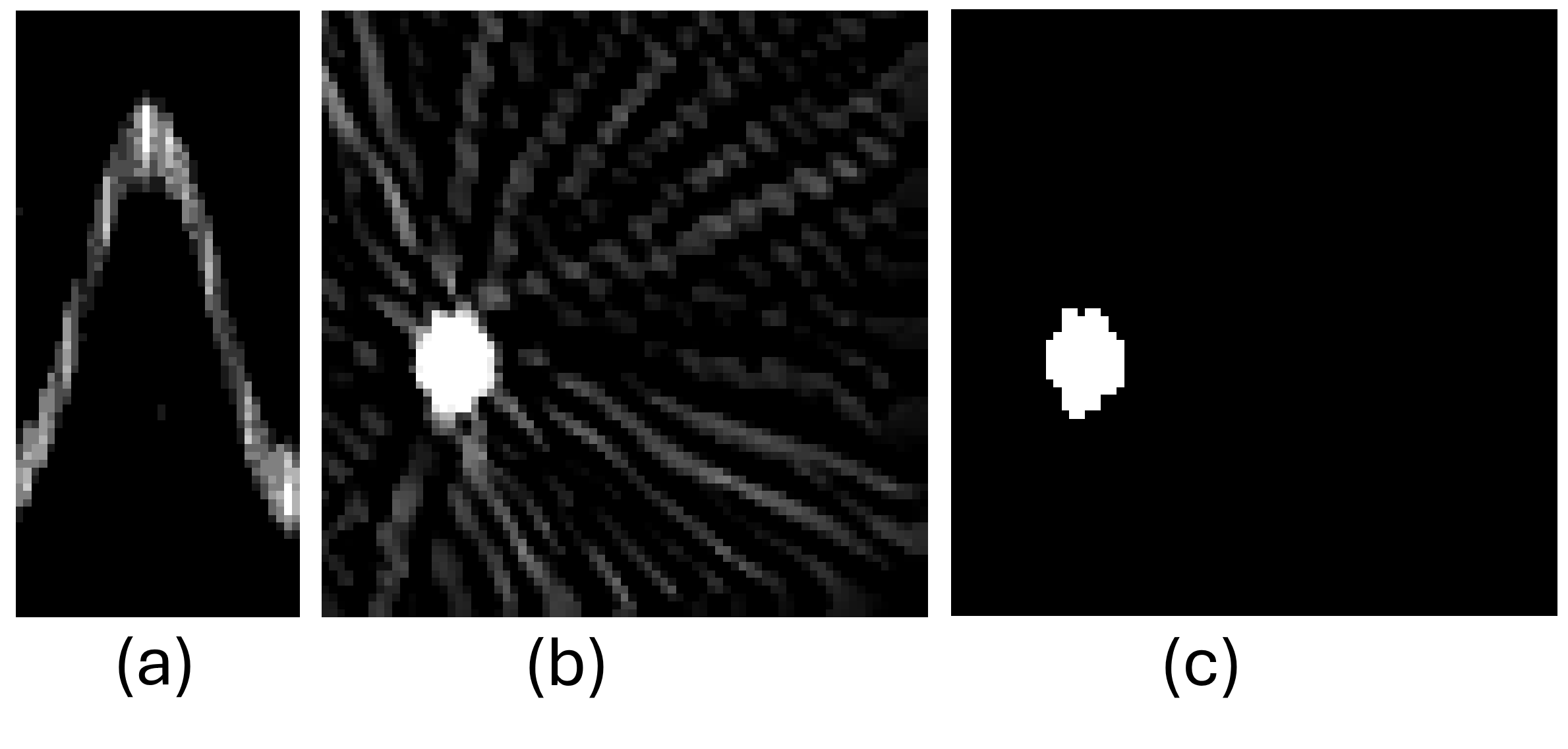} 
    \caption{Example illustrating the grain reconstruction process. (a) Sinogram showing the number of matched diffraction spots for this grain orientation at each translation step (vertical axis) and each rotation angle (horizontal axis). (b) Reconstructed grain using the sinogram in (a). (c) Segmented grain.}
    \label{fig:sinogram}
\end{figure}

Based on the indexed orientations, the number of diffraction spots matching each indexed orientation was determined using all individual, unmerged diffraction vectors obtained from each diffraction image at every scan position. This matching process involved comparing diffraction vectors, determined using the first ten \{hkl\} families, within a small deviation angle of 0.5$^\circ$ and a maximum distance of 10 pixels for the associated spots. The large number of \{hkl\} families were used to ensure the detection of the grains at more rotation angles, while the larger deviation angle was used to account for local variations in spot positions across translation steps. The number of matched diffraction spots was then used to create a sinogram for reconstructing the grain shape. An example of such a sinogram is shown in Fig.~\ref{fig:sinogram}(a). The sinograms were subsequently used to reconstruct grains using a filtered back-projection method. The reconstructed grain using the sinogram in Fig.~\ref{fig:sinogram}(a) is shown in Fig.~\ref{fig:sinogram}(b), based on which the grain shape was determined using a single-threshold segmentation approach, see Fig.~\ref{fig:sinogram}(c).

The grain map within the scanned layer was assembled by merging all the reconstructed grains. Overlapping pixels between grains were resolved by evaluating the pixel completeness values for all candidate grains at these pixels. The pixel completeness value was defined as the ratio of the observed number of spots from all images collected while the beam passed through the pixel and the theoretical number of spots expected for the candidate grain. Ultimately, each pixel was assigned to the grain with the highest completeness value. Additionally, unindexed pixels were assigned an orientation corresponding to the neighboring grain with the highest completeness value for the given pixels.

When the grain structure was reconstructed, the orientation of each grain can be further refined to reveal intragranular orientation variations by analyzing diffraction images collected as the beam passes through a given pixel, i.e., similar to the point-wise fitting in S3DXRD \cite{Hayashi2015,Hayashi2023}. To enhance orientation accuracy, more spots from up to ten \{hkl\} families were used, and a stricter angular deviation of 0.25$^\circ$ was applied for filtering diffraction spots. For this analysis, the actual pixel position was included in the refinement. 

\subsection{Validation}

To validate the orientations determined by Lab-3D$\mu$XRD, the LabDCT data was processed using GrainMapper3D, developed by Xnovo Technology ApS. A standard LabDCT spot segmentation routine was applied, incorporating rolling median background noise correction followed by a Laplacian of Gaussian-based filtering method to generate binary images for grain reconstruction. A fast geometric indexing algorithm \cite{Bachmann2019} was used to reconstruct the 3D volume of a total height of 400~$\mu$m. Reconstruction parameters were set to standard values, consistent with those used in \citeasnoun{Oddershede2022}. Grains were identified from neighboring regions with a maximum misorientation of 0.2$^\circ$.

The reconstructed 3D grains from synchrotron PCT reconstruction was downsampled by a factor of 3 to give the same effective voxel size as the absorption mask reconstruction of the LabDCT scan (i.e., 1.7 $\mu$m). The reconstructed PCT volume was then segmented into individual grains, providing shapes, sizes, and locations information but not their orientations. Hence, the PCT grains were registered to the Lab-3D$\mu$XRD results by minimizing the pairwise distances between the center-of-mass positions of the two maps.

\section{Results}
\subsection{Lab-3D\texorpdfstring{$\mu$}{mu}XRD Results}

\begin{figure}[ht]
    \centering
    \includegraphics[width=0.85\textwidth]{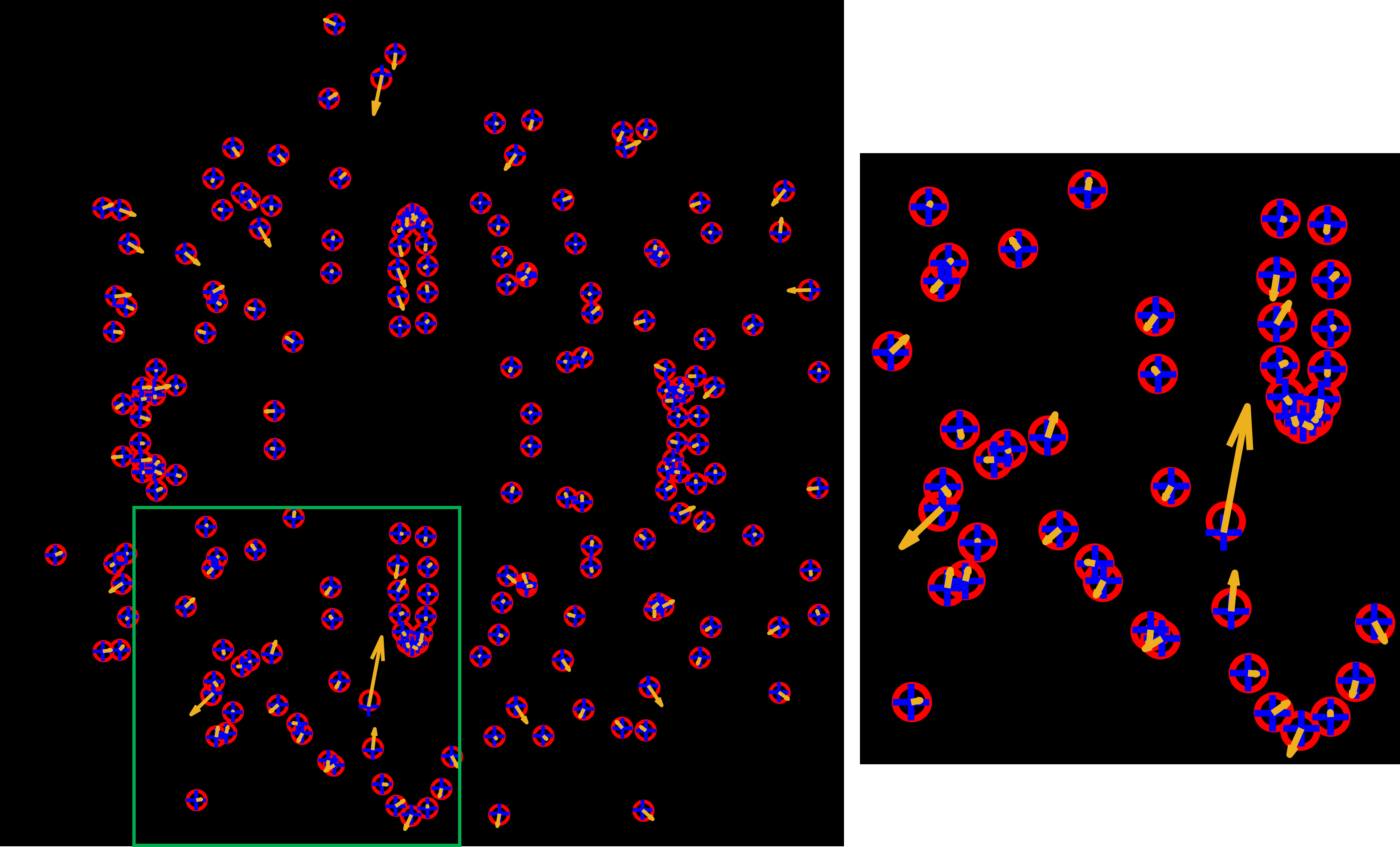} 
    \caption{Comparison between experimentally observed, merged diffraction spots (shown as red circles) and their theoretical positions (marked by blue crosses) for a relatively large grain. This image consolidates all the diffraction spots from the first ten \{hkl\} families observed in the 36 integrated diffraction images. The orange arrows indicate the direction and relative magnitude of the displacement between each pair of spots. The image on the right shows a magnified view of the region marked by the green rectangle.}
    \label{fig:diffraction_comparison}
\end{figure}

In total, 73 grain orientations were determined based on the LabDBB analysis of the Lab-3D$\mu$XRD data, among which one orientation did not result in satisfactory grain shape reconstructions. The completeness value for this orientation was 0.30 and it was therefore considered a false positive and excluded from further analysis. For the largest grain  observed within the scanned layer (with an equivalent circular diameter of (ECD) of 80~$\mu$m) approximately 100 and 250 diffraction spots were observed for the first three and ten \{hkl\} families, respectively (see Fig.~\ref{fig:diffraction_comparison}). In contrast, only 34 and 50 spots, respectively, are observed for the smallest grain (ECD of $\sim$10~$\mu$m). In all cases, a few falsely detected spots from noise may be included in the indexing; see, for example, those marked by the large orange arrows in Fig.~\ref{fig:diffraction_comparison}.

\begin{figure}[ht]
    \centering
    \includegraphics[width=0.85\textwidth]{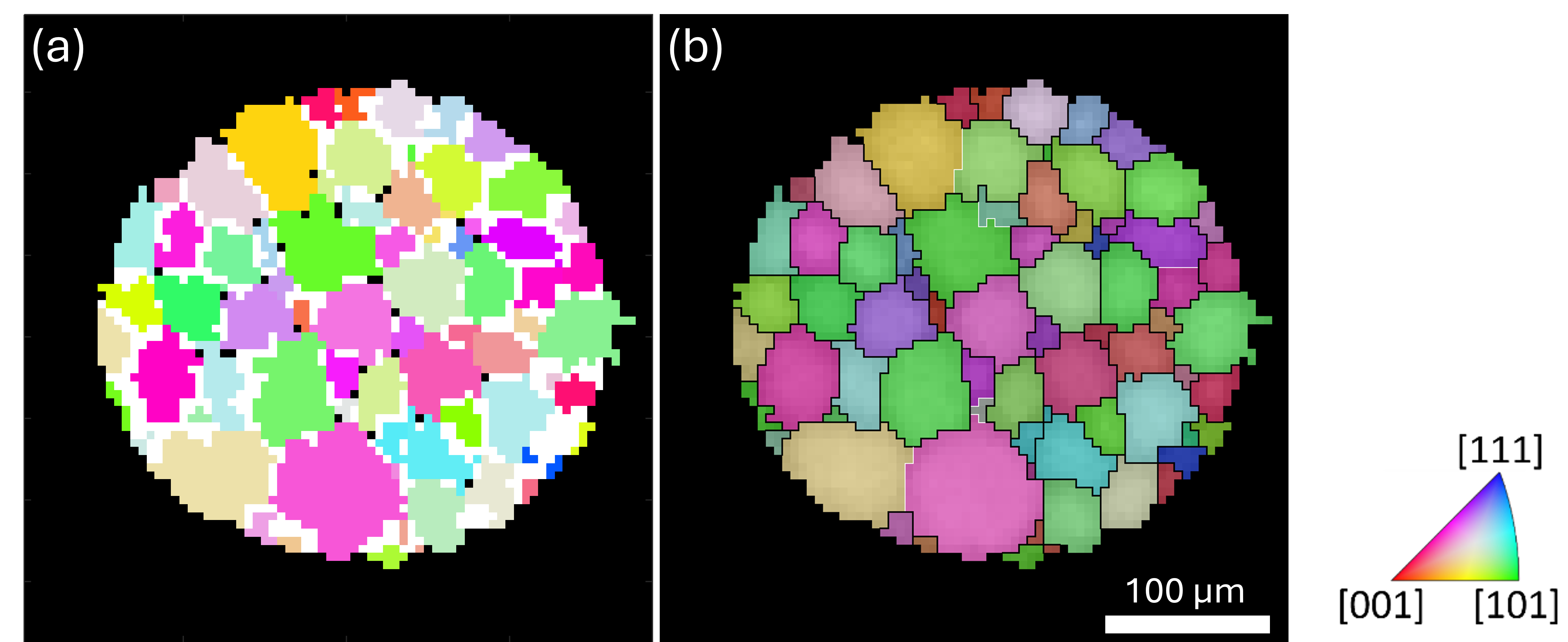} 
    \caption{Reconstructed grain map. (a) Segmented grains assembled directly. The white and black pixels inside the sample volume represent overlapped and unindexed pixels. (b) Grain map after processing. White and black lines indicate boundaries with misorientation angles larger than 2$^\circ$ and 15$^\circ$, respectively. The brightness of each pixel represents its completeness value.}
    \label{fig:grain_map}
\end{figure}

The raw reconstructed grain map is shown in Fig.~\ref{fig:grain_map}(a). A pixel size of 5~$\mu$m was used for the reconstruction, matching the scanning step size. Most of the overlapping and unindexed pixels are located at the grain boundaries, suggesting a representative reconstruction of the grains. However, two grains completely overlapped with other grains, and were eliminated during data cleaning. A cleaned map after resolving overlapped and unindexed pixels is shown in Fig.~\ref{fig:grain_map}(b), where the completeness value for each pixel is visualized by the brightness of the pixel. A total of 70 grains are present in the scanned layer, with most of the grain boundaries being high-angle boundaries.

\subsection{Comparison with LabDCT and PCT}

\begin{figure}[ht]
    \centering
    \includegraphics[width=0.8\textwidth]{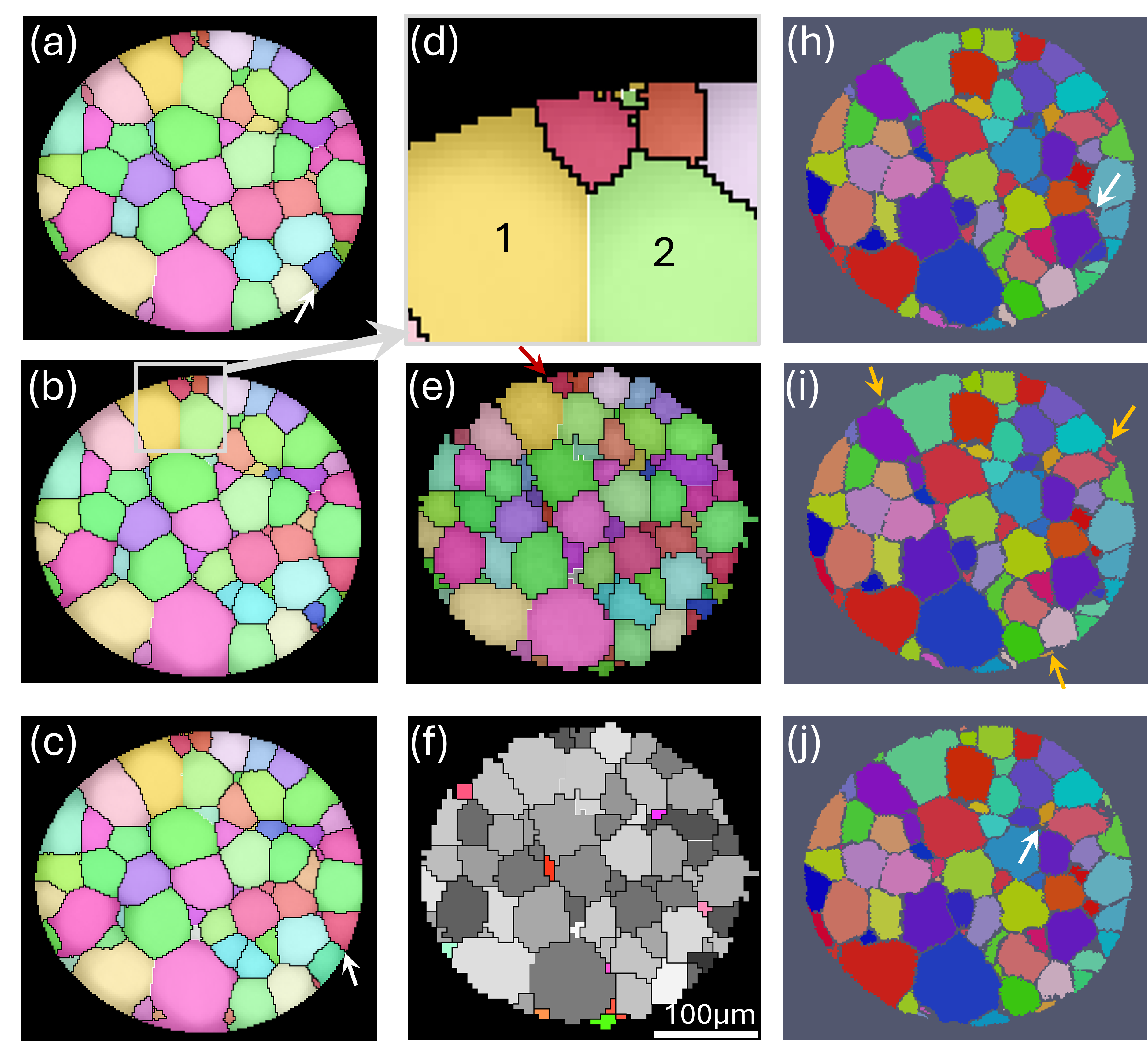} 
    \caption{Comparison among LabDCT, Lab-3D$\mu$XRD, and PCT. (a)–(c) Three sequential slices reconstructed using LabDCT with a voxel size of 2~$\mu$m. (d) A magnified view of the region marked by the gray rectangle in (b). (e) Lab-3D$\mu$XRD result reconstructed with a voxel size of 5~$\mu$m. (f) Grains indexed by Lab-3D$\mu$XRD but not LabDCT, highlighted in colors. (h)–(j) Three sequential slices showing the grain structure segmented based on PCT results with a voxel size of 1.7~$\mu$m. The grains in the PCT data are randomly colored.}
    \label{fig:comparison_slices}
\end{figure}

The Lab-3D$\mu$XRD results are compared with the LabDCT and PCT results in Fig.~\ref{fig:comparison_slices}. Three sequential slices from the scanned volume in both LabDCT and PCT were selected for a comprehensive comparison. A visual comparison of the three slices in the PCT data with the Lab-3D$\mu$XRD slice reveals that the middle slice (Fig.~\ref{fig:comparison_slices}(i)) provides the best match, as the other two slices contain missing grains, with two examples marked by the white arrows. A similar conclusion can be drawn for the LabDCT data, where the middle slice, Fig.~\ref{fig:comparison_slices}(b), provides the best match. It is important to note that since the missing grains in the two neighboring slices (at opposite sides) are close in $(x,y)$ coordinates, the mismatch between these slices and the Lab-3D$\mu$XRD slice cannot be attributed to sample volume tilt between different measurements. This is particularly true for the LabDCT data, as they were collected using the same sample mounting as Lab-3D$\mu$XRD.

Since the synchrotron PCT data was acquired with a finer resolution and absorption contrast is sensitive to the density difference of the alpha phase precipitated at the grain boundaries, the PCT segmented grain structure is considered the "ground truth" for the 3D grain shape in this study. A close comparison of individual grains suggests that Lab-3D$\mu$XRD was able to detect nearly all the grains present in the best-matched PCT slice, except for the three small grains marked by the orange arrows in Fig.~\ref{fig:comparison_slices}(i). These grains are located at the surface of the sample and are less than 7~$\mu$m in size on this slice. Compared to LabDCT, Lab-3D$\mu$XRD detects nine additional grains with sizes around 10–20~$\mu$m, as shown in Fig.~\ref{fig:comparison_slices}(f). For the remaining matched grains, the misorientation between the Lab-3D$\mu$XRD and LabDCT data is typically below 0.1$^\circ$, documenting the high precision in Lab-3D$\mu$XRD orientation determination.

In addition, several small grains or voxels indexed by LabDCT, such as those in the region marked by the gray rectangle in Fig.~\ref{fig:comparison_slices}(b) and magnified in Fig.~\ref{fig:comparison_slices}(d), are not indexed by Lab-3D$\mu$XRD. The orientations of these voxels match those of the two next-neighboring large grains (the yellowish grain \#1 and greenish \#2). Since these LabDCT grains are also absent in the corresponding PCT slice, they are considered false positives, likely resulting from the space-filling algorithm used during indexing. 

Finally, Lab-3D$\mu$XRD detects one small grain, marked by the red arrow in Fig.~\ref{fig:comparison_slices}(e), which is not present in the PCT data. Since this grain is also detected by LabDCT and the overall shape of this grain, together with the neighboring yellowish grain \#1, matches the corresponding large turquoise grain in the PCT data, it is considered correctly identified. The misorientation of the boundary between these two grains is 45.8$^\circ$⟨0.19, -0.08, 0.98⟩, which does not belong to any low-index CSL boundaries. It is unknown why it has not been properly detected by PCT.

\subsection{Intragranular orientation}
\begin{figure}[ht]
    \centering
    \includegraphics[width=0.85\textwidth]{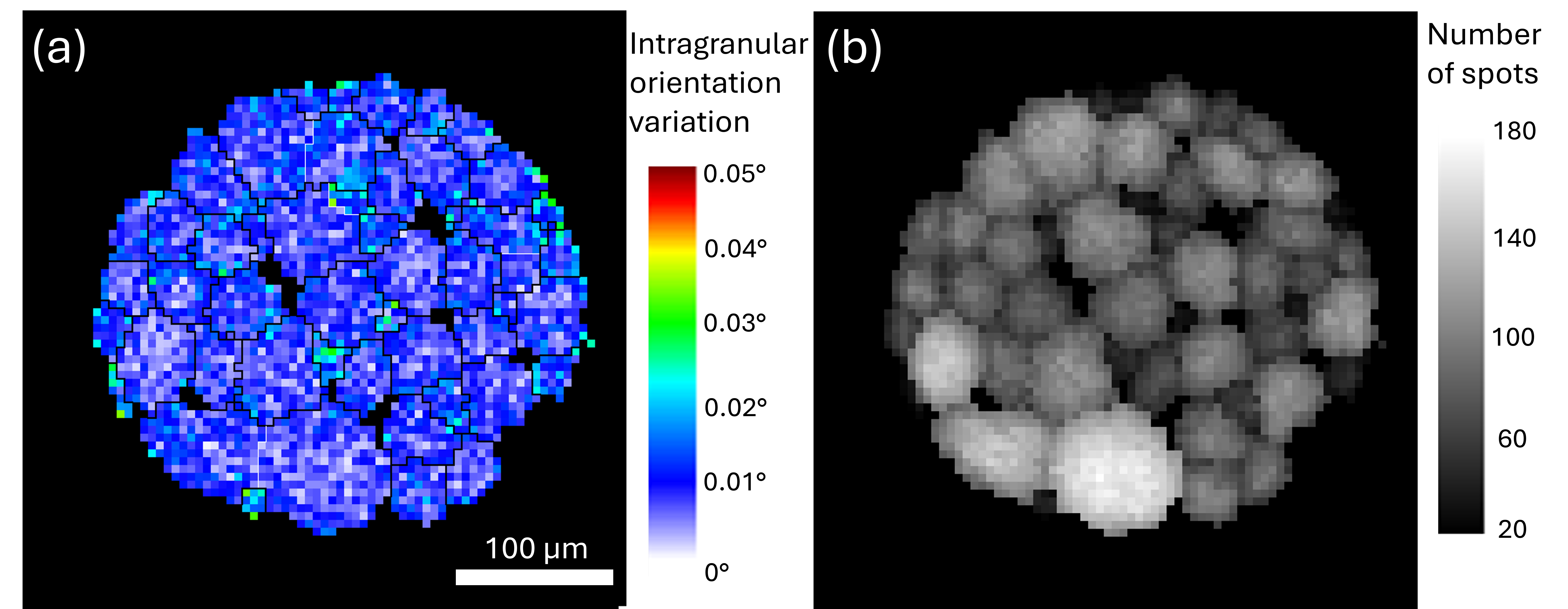} 
    \caption{(a) Intragranular orientation variation, calculated as the deviation from the average orientation of the grain, for the 50 largest grains. (b) Number of detected spots for each pixel.}
    \label{fig:intragranular_variation}
\end{figure}

The intragranular orientation quantified as the deviation angle of each pixel to the grain average orientation is shown in Fig.~\ref{fig:intragranular_variation}. Since the accuracy is directly related to the number of detected spots, only the 50 largest grains are shown here. A maximum deviation of $\sim$0.03$^\circ$ from the average orientation was detected in some grains, particularly among a few surface grains. Such small orientation changes may arise from surface damage of the sample (see, for example, the elongated spot in Fig.~\ref{fig:diffraction_spots}(d)) but could also result from noise, as they are most likely associated with a reduced number of detected spots close to grain boundaries (and in small interior grains). The average misorientation between neighboring pixels within individual grains is 0.006–0.008$^\circ$. Considering the thermomechanical processing history of the present sample with a very well-annealed grain structure, this level of orientation variation ($\sim$0.01$^\circ$) likely represents the orientation uncertainty of the present dataset. This low uncertainty is attributed to the small ratio of detector pixel size to detector distance, as well as the large number of diffraction spots involved in the setup.

\section{Discussion}

\subsection{Feasibility of Lab-3D\texorpdfstring{$\mu$}{mu}XRD}
The above analysis and comparison clearly demonstrate the feasibility of Lab-3D$\mu$XRD for 3D grain mapping. Specifically, the proposed rotation and scanning strategies can provide depth-resolved grain orientation information. 
With the white beam employed in Lab-3D$\mu$XRD, the number of rotation angles can be significantly reduced — by a factor of $\sim$10 compared to synchrotron S3DXRD \cite{Hayashi2015}, which uses a focused monochromatic beam. This is because a single Laue diffraction image can contain more than ten diffraction spots from the same grain, whereas in S3DXRD, there is rarely more than one spot per rotation step per grain. Additionally, in S3DXRD, a unique $(hkl)$ plane can produce only two spots during a full 360$^\circ$ rotation, whereas in Lab-3D$\mu$XRD, a unique $(hkl)$ plane can produce many more spots from different X-ray energies, depending on the number of rotation steps and the direction of the corresponding diffraction vector (see Fig.~\ref{fig:diffraction_comparison}). Furthermore, the continuous X-ray spectrum eliminates the need for sample rocking during data acquisition, simplifying the overall process.

Besides the Pt-coated twin paraboloidal capillary optics used in the present setup, other types of focusing optics, such as polycapillary optics \cite{Lynch2019} and multilayered supermirrors \cite{Joensen1994}, may also be used. In that case, the front beamstop, which blocks the central part of the beam, can be omitted, resulting in diffraction spots with a full-circle shape. However, to the best of the authors' knowledge, the present Pt-coated capillary optics outperform polycapillary optics in terms of intensity gain, focal spot size, beam achromatism, and the X-ray energy range of the focused beam \cite{Seret2023}. The practical implementation of multilayered supermirrors in a laboratory setting has yet to be demonstrated. 

In this first implementation, a sample rotation axis perpendicular to the incoming beam is chosen. While this is the most common setup for tomographic data acquisition, a rotation axis inclined at a non-perpendicular angle can also be used for implementing Lab-3D$\mu$XRD, as recently demonstrated by \citeasnoun{Kim2023}. This inclined rotation axis setup may be beneficial for improving the spatial resolution along the vertical direction. 

\subsection{Advantages of Lab-3D\texorpdfstring{$\mu$}{mu}XRD}

Compared to existing laboratory-based techniques, Lab-3D$\mu$XRD offers several distinct advantages. Lab-3D$\mu$XRD demonstrates a capability in detecting small grains ($<$20~$\mu$m), which is beneficial for gathering grain neighborhood and topology information, and thus for studying grain growth. This enhanced detection capability arises from the novel use of paraboloidal optics, which provides two key advantages: (1) significantly increased X-ray intensity \cite{Seret2023} and (2) a narrow ‘1D’ gauge volume/channel that precisely interrogates the microstructure. The enhanced X-ray intensity enables stronger diffraction signals for spot detection, while the focused 1D beam ensures that the diffraction signal is linearly proportional to the grain size ($R$). In contrast, LabDCT’s signal is proportional to the entire 3D volume ($R^3$). The more balanced spot intensities from grains of different sizes allow Lab-3D$\mu$XRD to operate with longer exposure times to resolve small grains without detector saturation from the large grains. Furthermore, since diffraction spots always originate from a local volume at the focal point, the total number of spots per image is typically low (around 20–30 in the present case). This minimizes spot overlap. Consequently, grain indexing and reconstruction can be relatively simple.

More importantly, Lab-3D$\mu$XRD can provide intragranular orientation information. Compared to the Laue focusing effect, micro-beam Laue diffraction is less sensitive to lattice defects, implying that Lab-3D$\mu$XRD is capable of studying plastic deformation. For example, previous results show that Lab-$\mu$XRD, equipped with a photon-counting detector (see more details below), can detect diffraction signals from additively manufactured AlSi10Mg alloys \cite{Zhang2024}, although the crystallographic orientation could not be determined from a single diffraction image. With Lab-3D$\mu$XRD, more diffraction spots would be captured during sample rotation, enabling the determination of the orientation field within grains. Such a study is planned for the near future.

\subsection{Future Lab-3D\texorpdfstring{$\mu$}{mu}XRD method development directions}

In principle, improved grain detection can enhance the accuracy of grain boundary positioning. A practical approach to achieving this is to develop a forward simulation-based reconstruction method, similar to LabDCT \cite{Bachmann2019}, capable of resolving the grain structure at a voxel size smaller than the selected translation step size (5~$\mu$m in this work). A recently demonstrated super-resolution strategy based on the superposition of X-ray beam trajectories \cite{Kim2023a} could also be explored as a potential enhancement to spatial resolution.

A significant issue of the present Lab-3D$\mu$XRD is the long data acquisition time needed for mapping a representative 3D volume. Although the current experimental setup, which utilizes a conventional flat-panel detector, has extended the detection limit to 10~$\mu$m, further improvement remains constrained by the detector’s sensitivity and dynamic range. Consequently, detecting even smaller grains ($<$10~$\mu$m) remains challenging with the present setup.

\begin{figure}[ht]
    \centering
    \includegraphics[width=0.85\textwidth]{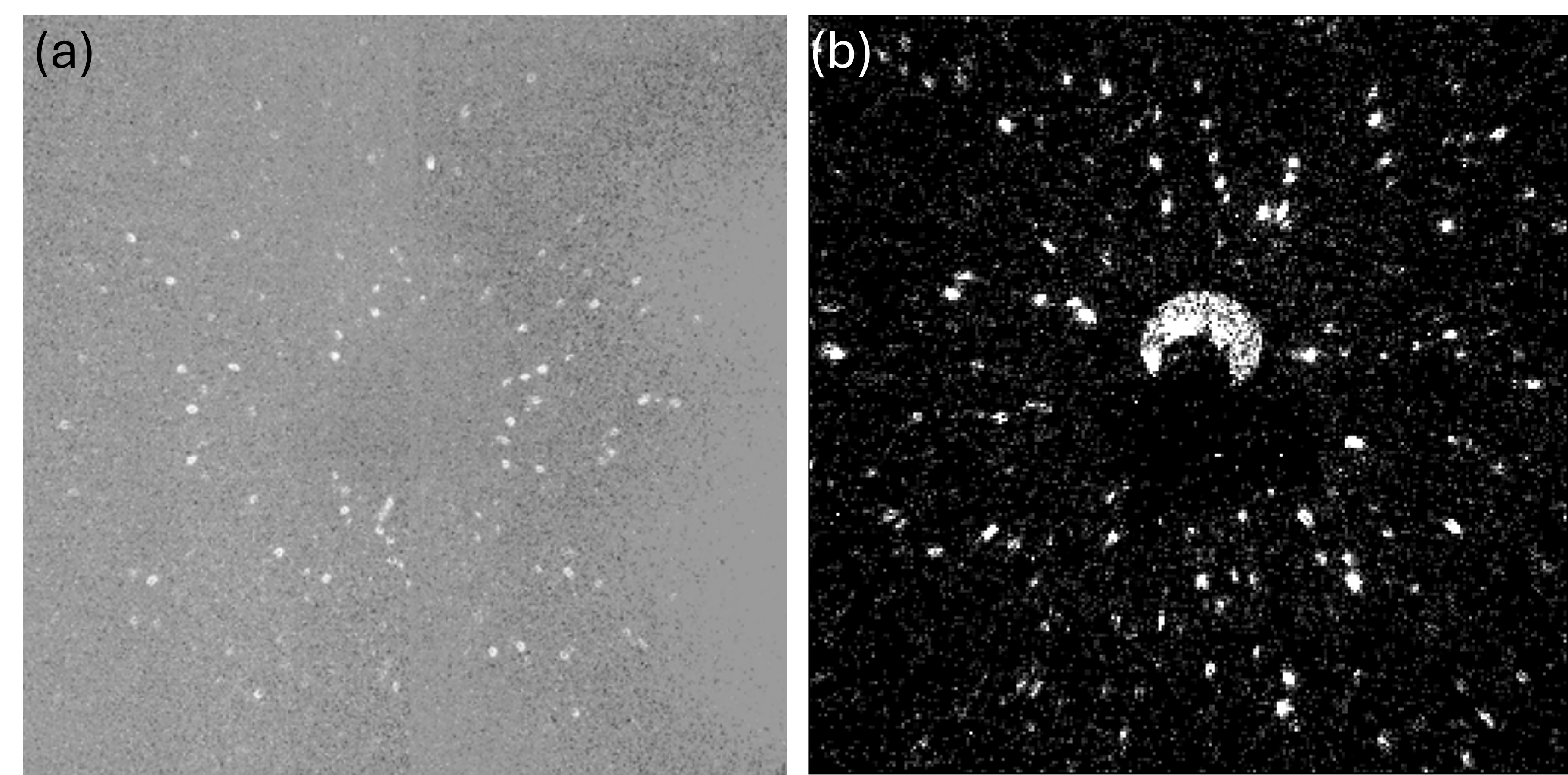} 
    \caption{Diffraction images collected from an iron powder sample (but not from the same location) using (a) the same flat panel detector with an exposure time of 170 s and (b) a photon-counting detector (Advacam ADVAPIX TPX3, 256$\times$256 pixels and 55~$\mu$m pixel size, placed at a $L_{sd}$ of ~14 mm) with an exposure time of 1 s. The intensities are after normalization using a reference image collected under the same condition but without the sample. }
    \label{fig:detector_performance}
\end{figure}

Further improvement of the efficiency of the optics, as suggested in ~\citeasnoun{Seret2023}, to eliminate both long-range and short-range defects of the mirror surface could reduce the focal spot size and increase the X-ray intensity. 
More importantly, the adoption of advanced photon-counting detectors \cite{Zhang2024}, which offer higher sensitivity, an improved dynamic range, as well as reduced noise, presents a viable solution to further decrease acquisition time and enhance the detection of smaller grains. A preliminary test on a pure iron sample has demonstrated that a realistic gain factor of $\sim$200 in exposure time is achievable with an Advacam ADVAPIX TPX3 photon-counting detector, as illustrated in Fig.~\ref{fig:detector_performance}. The sample was prepared from a powder with a nominal particle size of 5-8~µm and sintered at 850~°C for 0.5~hours in a borosilicate glass tube with inner and outer diameters of 80 and 100~µm, respectively, resulting in a relative density of about 75\%. The exact grain size after sintering is unknown. However, based on the high number of detected diffraction spots (about 80), it is likely that more than 10 grains were present within the illuminated volume channel. With increased exposure time, it is expected that the high photon sensitivity of the detector will enable the detection of grains as small as ~5~µm. By combining this enhanced detector performance with a 10–100$\times$ gain in X-ray flux—achievable using, for example, an intense X-ray source from a liquid metal jet anode \cite{David2004}, linear accumulation X-ray sources \cite{Yun2017} or tabletop synchrotron light source \cite{Yamada2014}—the exposure time could be reduced by an order of magnitude or more. Altogether, this advancement could enable Lab-3D$\mu$XRD to achieve 3D mapping of grains as small as 2-5~$\mu$m and of samples with a high defect content, such as additively manufactured materials \cite{Zhang2024} in laboratory settings. This would open a whole new arena for in-house 3D characterization.

Besides the hardware and software improvements, further optimization of experimental parameters, such as the maximum X-ray energy and power, scanning step size, number of rotation angles, and detector distance, is necessary. Typically, increasing the input power can enlarge the X-ray source size, which in turn leads to an increase in focal spot size. However, this enlargement may not significantly compromise the ring shape or symmetry of the focused beam with the present focusing optics. As a result, its impact on the diffraction spot shape could remain minimal. To quantify this effect, a ray-tracing simulation, such as a Monte Carlo method \cite{Bergback2013}, could be conducted. Moreover, the feasibility of the technique for other material systems, such as Al and steel, with varying sample sizes, absorption rates, and defect contents, should be explored.

Furthermore, the present proposed method can be extended or integrated into advanced systems to further enhance its capabilities. For example, the principles of Lab-3D$\mu$XRD can be adapted for use with synchrotron sources. This approach could simplify synchrotron-based setups by eliminating the need for complex designs and integration of differential aperture systems while maintaining the ability to perform high-resolution orientation and strain mapping. Additionally, adapting this method to synchrotrons would leverage the inherently higher flux of synchrotron X-ray beams, further improving the detection of smaller grains and intragranular features. At the same time, Lab-3D$\mu$XRD can be combined with existing techniques such as LabDCT and $\mu$CT, enabling multiscale and multimodal characterization of materials \cite{Zhang2024a,Knipschildt-Okkels2025}, significantly expanding the scope of materials research accessible to lab-scale facilities.

\section{Conclusions}

Lab-3D$\mu$XRD has been successfully implemented into a conventional X-ray CT system using Pt-coated twin paraboloidal capillary X-ray focusing optics, and in-house developed data acquisition and processing software. By comparing with LabDCT and synchrotron PCT, it is documented that grains in the size range of 10–20~$\mu$m can be well reconstructed by Lab-3D$\mu$XRD with an orientation uncertainty as low as $\sim$0.01$^\circ$. This capability provides excellent grain neighborhood information, albeit with poor temporal resolution. With commercially available advanced photon-counting detectors and more intense X-ray sources, Lab-3D$\mu$XRD is expected to become capable of detecting grains as small as 2–5~$\mu$m and of characterizing deformed materials. Since Lab-3D$\mu$XRD shares most of its hardware components with LabDCT, integrating both techniques within a single X-ray CT system is feasible, enabling comprehensive multiscale and multimodal characterization for gaining comprehensive insights into material behavior and properties.

\begin{acknowledgements}
Y Zhang and D Juul Jensen thank Villum Fonden, Denmark (grant MicroAM VIL 54495) for supporting this work. A Seret acknowledges the support from the Villum Experiment grant, project number 00028354. Dr. T. M. Ræder is acknowledged for his kind assistance in creating Fig. 1 and for his valuable comments on the manuscript.

\end{acknowledgements}

\begin{funding}
Villum Fonden, D Juul Jensen, MicroAM VIL 54495; Villum Fonden, Y Zhang, 00028354.
\end{funding}

\ConflictsOfInterest{The authors have not conflicts of interest.}

\bibliography{iucr} 

\clearpage
\appendix
\renewcommand{\thefigure}{S\arabic{figure}} 
\setcounter{figure}{0} 



\end{document}